\documentclass[conference, a4paper]{IEEEtran}

\ifCLASSINFOpdf
\else
\fi

\usepackage{array}
\usepackage{fixltx2e}
\usepackage{url}

\usepackage[dvipdfmx]{graphicx}
\usepackage{diagbox}

\begin{document}

\title{\it Polysemy Detection in Distributed Representation of Word Sense}

\author{\IEEEauthorblockN{Kana Oomoto\IEEEauthorrefmark{1},
Haruka Oikawa\IEEEauthorrefmark{1},
Eiko Yamamoto\IEEEauthorrefmark{2},
Mitsuo Yoshida\IEEEauthorrefmark{1},
Masayuki Okabe\IEEEauthorrefmark{3} and
Kyoji Umemura\IEEEauthorrefmark{1}}
\IEEEauthorblockA{\IEEEauthorrefmark{1}Department of Computer Science and Engineering,
Toyohashi University of Technology,
Toyohashi, Japan\\
k133309@edu.tut.ac.jp, h153312@edu.tut.ac.jp, yoshida@cs.tut.ac.jp, umemura@tut.jp
}
\IEEEauthorblockA{\IEEEauthorrefmark{2}Faculty of Economics and Information,
Gifu Shotoku Gakuen University,
Gifu, Japan\\\
eiko@gifu.shotoku.ac.jp
}
\IEEEauthorblockA{\IEEEauthorrefmark{3}Faculty of Management and Information Systems,
Prefectural University of Hiroshima,
Hiroshima, Japan\\
okabe@pu-hiroshima.ac.jp}
}

\IEEEoverridecommandlockouts
\IEEEpubid{\makebox[\columnwidth]{978-1-4673-9077-4/17/\$31.00~\copyright~2017 IEEE \hfill} \hspace{\columnsep}\makebox[\columnwidth]{}} 

\maketitle

\begin{abstract}
In this paper, we propose a statistical test to determine whether a given word is used as a polysemic word or not.
The statistic of the word in this test roughly corresponds to the fluctuation in the senses of the neighboring words and the word itself.
Even though the sense of a word corresponds to a single vector, we discuss how polysemy of the words affects the position of vectors.
Finally, we also explain the method to detect this effect.
\end{abstract}

\renewcommand\IEEEkeywordsname{Keywords}

\IEEEpeerreviewmaketitle

\section{Introduction}
Distributed representation of word sense provides us with the ability to perform several operations on the word.
One of the most important operations on a word is to obtain the set of words whose meaning is similar to the word, or whose usage in text is similar to the word.
We call this set the neighbor of the word.
When a word has several senses, it is called a polysemic word.
When a word has only one sense, it is called a monosemic word.
We have observed that the neighbor of a polysemic word consists of words that resemble the primary sense of the polysemic word.
We can explain this fact as follows.
Even though a word may be a polysemic, it usually corresponds to a single vector in distributed representation.
This vector is primarily determined by the major sense, which is most frequently used.
The information about a word's minor sense is subtle, and the effect of a minor sense is difficult to distinguish from statistical fluctuation.

To measure the effect of a minor sense, this paper proposes to use the concept of {\it surrounding uniformity}.
The surrounding uniformity roughly corresponds to statistical fluctuation in the vectors that correspond to the words in the neighbor.
We have found that there is a difference in the surrounding uniformity between a monosemic word and a polysemic word.
This paper describes how to compute surrounding uniformity for a given word, and discuss the relationship between surrounding uniformity and polysemy.

\section{Related Work}
The distributed word representation can be computed as weight vectors of neurons, which learn language modeling ~\cite{Bengio:2003:NPL:944919.944966}.
We can obtain a distributed representation of a word using the Word2Vec software~\cite{NIPS2013_5021} which enable us to perform vector addition/subtraction on a word's meaning.
The theoretical background is analyzed by~\cite{NIPS2014_5477}, where the operation is to factorize a word-context matrix, where the elements in the matrix are some function of the given word and its context pairs.
This analysis gives us insight into how the vector is affected by multiple senses or multiple context sets.
If a word has two senses, the obtained representation for the word will be a linearly interpolated point between the two points of their senses.

The importance of multiple senses is well recognized in word sense detection in distributed representation.
The usual approach is to compute the corresponding vectors for each sense of a word~\cite{Reisinger:2010:MVM:1857999.1858012,Huang:2012:IWR:2390524.2390645}.
In this approach, first, the context is clustered.
Then, the vector for each cluster is computed.
However, the major problem faced by this approach is that all target words need to be assumed as polysemic words first, and their contexts are always required to be clustered.
Another approach is to use external language resources for word sense, and to classify the context~\cite{chen-liu-sun:2014:EMNLP2014}.
The problem with this approach is that it requires language resources of meanings to obtain the meaning of a polysemic word.
If we know whether a given word is polysemic or monosemic thorough a relatively simple method, we can concentrate our attention on polysemic words.

\section{Senses and Contexts}
In this paper, we assume that the sense of a word is determined by the distribution of contexts in which the word appears in a given corpus.
If a word comes to be used in new contexts, the word comes to have a new sense.
If we could have an infinitely sizes corpus, this sense might converge into the sense in the dictionary.
In reality, the size of the corpus in hand is limited, and some senses indicated in a dictionary may not appear in the corpus.
The distinction between the senses in a dictionary and the senses in the corpus is important in this paper, because it is crucial for discussing polysemy.
All discussions in this paper depend on the corpus in hand.
We now use the FIL9 corpus (http://mattmahoney.net/dc/textdata), which primarily consists of a description of believed facts, rather than conversations.
We can expect that the senses that are mainly used in conversation would not appear in this corpus.

In this paper, we analyze auxiliary verbs, which are polysemic words from a dictionary.
If the corpus is limited to a description of believed facts, we may regard auxiliary verbs as monosemic words, since their contexts are limited.
In addition, we particularly analyze the relationship between the auxiliary verb ``may'', and name of the month ``May''.
In the dictionary, these two are regarded as two different words, rather than as two different senses of one word.
By ignoring the upper/lower case characters, these two words have same character sequence and the word ``may'' becomes a polysemic word, which has two types of context in the given corpus.

\section{Proposed Method}
Our proposed method is based on the following measures.
Let $\vec{w}$ be the vector corresponding to the given word.
Let $N$ be the size of the neighbor, such as $4$.
First, we choose $N$ neighboring words whose angle with the given word is the smallest.
This operation is already implemented in the Word2Vec software.
Let $\vec{a_i}$($\vec{w}$) be the vectors corresponding to $i$th vector of the neighbor of the word.

We choose the uniformity of vectors, which can be regarded as general case of triangle inequality.
The uniformity of a set of vectors is a ratio, i.e., the size of the vector of the vector addition of the vectors divided by the scalar sum of the sizes of the vectors.
If and only if all directions of the vectors are the same, the uniformity becomes 1.0.
We compute this uniformity for the neighbors, including the word itself.
Surrounding Uniformity (SU) can be expressed as follows:\\
$$SU(\vec{w}) = \frac{|\vec{s}(\vec{w})|}{|\vec{w}| + \sum_{i}^{N}|\vec{a_i}(\vec{w})|}$$
where $$\vec{s}(\vec{w}) = \vec{w} + \sum_{i}^{N} \vec{a_i}(\vec{w}).$$\\

When computing SU, we consider the set of words whose vectors are reliable. 
We choose these words as the most frequently appearing words in corpus. 
The size of words is denoted as $limit$. 
If a word is not in this set, or the word does not have sufficient number of neighbors in this set, we consider that the value of SU is undefined, and that the word does not have this value.

Our method performs a statistical test to determine whether a given word is used polysemously in the text, according to the following steps: 
\begin{enumerate}
\item Setting $N$, the size of the neighbor.
\item Choosing $N$ neighboring words $a_i$ in the order whose angle with the vector of the given word $w$ is the smallest.
\item Computing the surrounding uniformity for $a_i$($0<i \leq N$) and $w$.
\item Computing the mean $m$ and the sample variance $\sigma$ for the uniformities of $a_i$.
\item Checking whether the uniformity of $w$ is less than $m-3\sigma$.
If the value is less than $m-3\sigma$, we may regard $w$ as a polysemic word.
\end{enumerate}
This is a basic statistical test~\cite{Wasserman200410} to detect outliers. 

Note that we cannot compute the variance if some $a_i$ does not have the value of SU.
Further, it may be also possible that all $a_i$ may have the same SU, sharing identical neighbors.
In this case, the variance becomes an extreme value, that is, $0$. 
In these cases, we consider that we cannot perform the statistical test.

\section{Experimental Settings and Examples of Calculation}
We used FIL9, which is freely available as the test corpus for Word2Vec and is derived from Wikipedia. 
We compute $200$-dimensional distributed vector representations with default parameter.
In this situation, all-uppercase are converted into lower case. 
This is why all proper nouns are in lower case in this example. 
First we selected stable words as the 1000 words that appear most frequently in the text. 
We compute surrounding uniformity of these words. 
We define the given word $w$ and its neighboring word $a_i$ are limited to stable words. 
We then determine the search scope for stable neighboring words and set $N$, which is the number of neighbors used to compute the surrounding uniformity, to $4$. 
For example, if there are 7 stable words in the search scope, we use only the top 4 words to compute the surrounding uniformity.

Table~\ref{tab:auxiliary_verb} shows the uniformity of auxiliary verbs in this setting. 
We were able to compute the surrounding uniformity for 160 words; for the remaining 840 words, there were fewer than the required $4$ stable neighboring words in the search scope and the surrounding uniformity could not be determined. 

\begin{table*}[tbp]
\renewcommand{\arraystretch}{1.3}
\centering
\caption{
Auxiliary verbs, their neighboring words, and surrounding uniformities.
The neighboring words of an auxiliary verb consist of other auxiliary verbs.
The word ``may'' has a small surrounding uniformity, although its neighboring words consist of auxiliary verbs.
}
\label{tab:auxiliary_verb}
\begin{tabular}{|l|llll|r||l|}
\hline 
\bfseries word ($w$) & \multicolumn{4}{|l|}{\bfseries neighboring words ($a_i$, $N=4$)} & \bfseries surrounding uniformity & \bfseries test \\
\hline \hline
could &  would &  will &  might &  must & 0.9290 & -\\
will &  would &  must &  could &  should & 0.9266 & -\\
would &  could &  will &  might &  should & 0.9266 & -\\
must &  cannot &  will &  should &  could & 0.9253 & -\\
can &  cannot &  must &  could &  will & 0.9252 & -\\
should &  must &  could &  might &  will & 0.9232 & -\\
cannot &  must &  can &  could &  might & 0.9221 & -\\
might &  would &  could &  should &  cannot & 0.9179 & -\\
may &  can &  should &  might &  will & 0.8917 & significant\\
\hline
\end{tabular}
\end{table*}

For the case of the word ``may'', neighbor words are ``can'', ``should'', ``might'', and ``will''.
Their surrounding uniformities are, 0.9252 (``can''), 0.9232 (``should''), 0.9179 (``might''), and 0.9266 (``will'').
Then $m$ is equal to 0.9232, and $\sigma$ is equal to 0.0038.
Therefore, $m-3\sigma$ is 0.9118, which is greater than 0.8917 (``may'').
Since the surrounding uniformity of the word ``may'' is regarded as an outlier, we think of ``may'' as polysemic.
In this setting, the word ``may'' is polysemic because the program works in a case-insensitive mode, and the word ``may'' could be both an auxiliary verb and the name of a month.

The next example is the word ``might'', whose surrounding uniformity is smaller than every neighbor word.
For the word ``might'', neighbor words are ``would'', ``could'', ``should'', and ``cannot''.
Their surrounding uniformities are 0.9266 (``would''), 0.9290 (``could''), 0.9232 (``should''), and 0.9224 (``cannot'').
Hence, $m$ is equal to 0.9253, and $\sigma$ is equal to 0.0032.
Therefore, $m-3\sigma$ is 0.9157, which is less than 0.9179 (``might'').
We cannot say 0.9179 is an outlier, and thus we cannot say the word ``might'' is polysemic.

\begin{figure*}[tbp]
\centering
\includegraphics[width=4.2in]{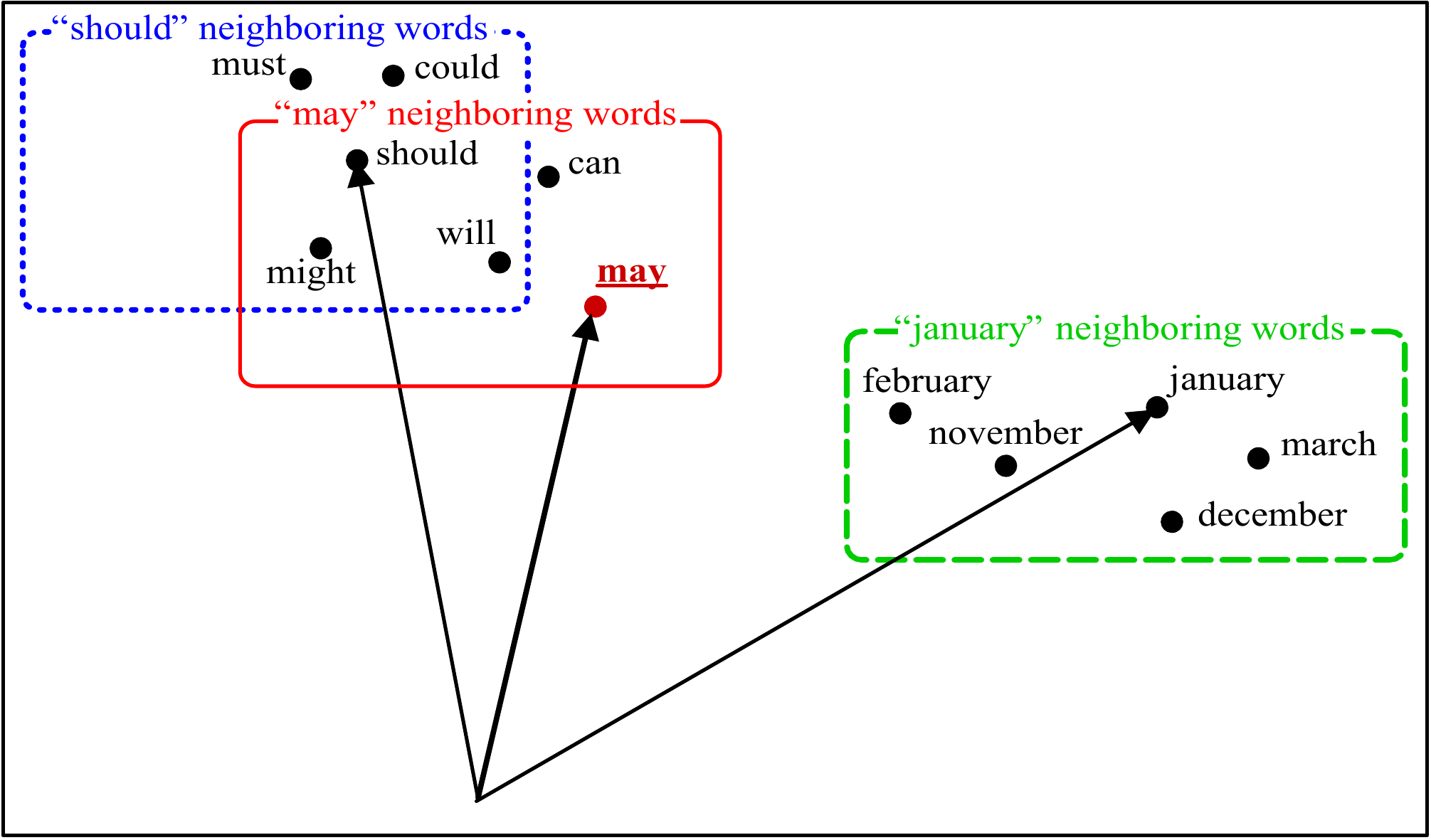}
\caption{
Relation of the words: ``should'', ``may'', and ``january''.
The distributed representation of the word ``may'' is placed near other auxiliary verbs.
Regarding the linear nature of Word2Vec, it is natural that the position should be between auxiliary verbs and the names of the months.
}
\label{fig:map}
\end{figure*}

Figure~\ref{fig:map} shows the distribution of vectors.
The vector of ``may'' is placed in the interpolated position between ``may'' as an auxiliary verb and ``may'' as the name of a month.
Since the word ``may'' is more frequently used as auxiliary verb, the vector is placed near other auxiliary verbs.
However, the position of ``may'' could be an outlier for other auxiliary verbs.

In addition, we should show the results of names of months because these names will have the same contexts when the word is used as the name of a month.
The word ``may'' has other contexts as auxiliary verbs.
The word ``august'' has the sense of an adjective in the dictionary.
The word ``march'' has a sense of a verb.
Other names are monosemic words in the dictionary.
Table~\ref{tab:month} shows the surrounding uniformity for all the names of the months.
\begin{table*}[tbp]
\renewcommand{\arraystretch}{1.3}
\centering
\caption{
Names of the months, their neighboring words, and surrounding uniformities.
Only ``may'', which has the smallest surrounding uniformity, pass the statistical test.
Although the word ``may'' might be used as the name of a month, the corresponding vector is near the auxiliary verbs.
}
\label{tab:month}
\begin{tabular}{|l|llll|r||l|}
\hline 
\bfseries word ($w$) & \multicolumn{4}{|l|}{\bfseries neighboring words ($a_i$, $N=4$)} & \bfseries surrounding uniformity & \bfseries test \\
\hline \hline
december &  november &  october &  march &  september & 0.9817 & -\\
january &  february &  december &  november &  march & 0.9816 & -\\
october &  november &  december &  february &  september & 0.9815 & -\\
june &  july &  march &  december &  april & 0.9814 & -\\
april &  march &  december &  september &  july & 0.9810 & -\\
november &  december &  october &  september &  january & 0.9810 & -\\
february &  january &  december &  march &  october & 0.9809 & -\\
march &  december &  april &  june &  february & 0.9806 & -\\
september &  december &  august &  april &  november & 0.9804 & -\\
july &  june &  december &  august &  september & 0.9804 & -\\
august &  september &  july &  june &  april & 0.9802 & -\\
may &  can &  should &  might &  will & 0.8917 & significant\\
\hline
\end{tabular}
\end{table*}

If we apply the test, only the word ``may'' passes the test.
The example that fails the test is the word ``august'', whose surrounding uniformity is also smaller than every neighbor word.
For the case of the word ``august'', $m$ is equal to 0.9808, and $\sigma$ is equal to 0.0005.
Therefore, $m-3\sigma$ becomes 0.9793, which is less than 0.9802 (``august'').
We cannot say the word ``august'' is polysemic, but the value of uniformity is very close to the lower bound.
Other names have a greater uniformity than the corresponding lower bound.
In summary, the proposed method can detect the polysemic ``may'', but cannot detect the polysemicity of ``august'' and ``march''.

Although we can claim nothing if the statistical test fails, even the negatives have a practical value for this test.
For the case of the word ``august'', it can be used as an adjective.
Although we cannot say the word ``august'' is polysemic from the proposed procedure, we cannot claim that the word ``august'' is monosemic.
We think this failure is caused by a few, if any, contexts of ``august'' as an adjective.
In that case, the clustering context will be difficult in practice.
Therefore, the proposed test will be meaningful even for a negative result, when the result is used to judge whether further analysis of the context is worthwhile.
This discussion should be also true for the word ``march'', which may be used as a verb.
 
There are other interesting words for which the proposed method detects polysemicity.
These words are ``james'', ``mark'', and ``bill''.
The neighboring words are names of persons, such as ``john'', ``richard'', ``robert'', ``william'', ``david'', ``charles'', ``henry'', ``thomas'', ``michael'', and ``edward''.
``mark'' and ``bill'' have the same spell of the regular noun.
The word ``james'' does not have such words and is subject to error analysis. 

\section{Evaluation}
First, we set the value of $limit$ to $1000$, and $N$ to $4$.
We then performed the statistical test of these 1000 words.
From these, 33 words passed test, and we assume that these words belong to the set POLY.
Further, we are unable to performs the statistical test for 127 words.
We say that the remaining 840 words belong to the set MONO.

As evaluation, we attempted to measure the agreement of human judgment for the all words of POLY and MONO.
However, during the valuation, we found that many of the errors come from the problem of Word2Vec.
For example, the vector of ``sir'' and the vector of ``william'' are very close because ``sir william'' should be very close to ``william''. 
This is similar for ``w'' and ``george".

Therefore, we first selected words whose 10 neighboring words seem reasonable neighbors for human judgments, and performed human judgments of polysemicity.
We also focused the words that have bigger SU than 0.75. 
This is because the statistical test will be reliable when SU is large. 
Table~\ref{tab:SU_poly} shows that list of words that passed the test, and have higher SU than 0.75. 
\begin{table*}[tbp]
\renewcommand{\arraystretch}{1.3}
\centering
\caption{
Evaluated Words and its neighbor that passes the statistical test.
}
\label{tab:SU_poly}
\begin{tabular}{|l|rrrr|l||l|l|}
\hline 
\bfseries word ($w$) & \multicolumn{4}{|l|}{\bfseries neighboring words and neighboring word's SU ($a_i$, $N=4$)} & \bfseries SU & \bfseries computer & \bfseries human \\
\hline \hline
	may & can (0.9252) & should (0.9232) & might (0.9179) & will (0.9266) & 0.8917 & poly & poly\\
	james & john (0.9067) & robert (0.8984) & richard (0.8984) & william (0.8950) & 0.8788 & poly & mono \\
	mark & peter (0.8675) & michael (0.8881) & david (0.8945) & smith (0.8702) & 0.8249 & poly & poly\\
	bill & peter (0.8675) & david (0.8945) & michael (0.8881) & richard (0.8984) & 0.8196 & poly & poly\\
\hline
\end{tabular}
\end{table*}
Table~\ref{tab:SU_poly} shows all the words in POLY that are judged by human. 
Similarly Table~\ref{tab:SU_mono} shows all the words in MONO that are judged by human. 
\begin{table*}[tbp]
\renewcommand{\arraystretch}{1.3}
\centering
\caption{
Evaluated Words that does not pass the statistical test.
}
\label{tab:SU_mono}
\begin{tabular}{|l|rrrr|l||l|l|}
\hline 
\bfseries word ($w$) & \multicolumn{4}{|l|}{\bfseries neighboring words and neighboring word's SU ($a_i$, $N=4$)} & \bfseries SU & \bfseries computer & \bfseries human \\
\hline \hline
	december & november (0.9810) & october (0.9815) & march (0.9806) & september (0.9804) & 0.9817 & mono & mono\\
	november & december (0.9817) & october (0.9815) & september (0.9804) & january (0.9816) & 0.9810 & mono & mono\\
	august & september (0.9804) & july (0.9804) & june (0.9814) & april (0.9810) & 0.9802 & mono & mono\\
	three & four (0.9730) & five (0.9730) & six (0.9737) & seven (0.9737) & 0.9730 & mono & mono\\
	will & would (0.9266) & must (0.9253) & could (0.9290) & should (0.9232) & 0.9266 & mono & mono\\
	two & three (0.9730) & five (0.9730) & four (0.9730) & zero (0.9221) & 0.9221 & mono & mono\\
	richard & robert (0.8984) & william (0.8950) & john (0.9067) & david (0.8945) & 0.8984 & mono & mono\\
	henry & edward (0.8848) & william (0.8950) & charles (0.8922) & richard (0.8984) & 0.8922 & mono & mono\\
	tv & television (0.8293) & radio (0.8037) & shows (0.7953) & network (0.7835) & 0.8293 & mono & poly\\
	make & give (0.8150) & find (0.8465) & get (0.8465) & makes (0.7517) & 0.8201 & mono & mono\\
	pennsylvania & ohio (0.8295) & michigan (0.8336) & virginia (0.7748) & township (0.8171) & 0.8186 & mono & mono\\
	french & italian (0.8116) & dutch (0.8132) & german (0.8128) & english (0.7765) & 0.8177 & mono & mono\\
	spanish & french (0.8177) & english (0.7765) & dutch (0.8132) & italian (0.8116) & 0.8164 & mono & mono\\
	dutch & french (0.8177) & german (0.8128) & spanish (0.8164) & english (0.7765) & 0.8132 & mono & mono\\
	go & get (0.8465) & come (0.7838) & take (0.7892) & move (0.8010) & 0.8116 & mono & mono\\
	radio & television (0.8293) & tv (0.8293) & network (0.7835) & news (0.7342) & 0.8037 & mono & mono\\
	move & get (0.8465) & turn (0.7800) & go (0.8116) & find (0.8465) & 0.8010 & mono & mono\\
	founded & established (0.7967) & formed (0.7899) & introduced (0.7835) & built (0.7464) & 0.7967 & mono & mono\\
	child & daughter (0.8420) & woman (0.7672) & mother (0.8796) & children (0.7246) & 0.7829 & mono & mono\\
	studies & research (0.7810) & study (0.7798) & education (0.7542) & philosophy (0.7632) & 0.7798 & mono & mono\\
\hline
\end{tabular}
\end{table*}
We have sampled words from MONO because there are many words in MONO.
In these tables, the SU of surrounding words are also presented.

Table~\ref{tab:count_human_total} shows the confusion matrix for computer human judgment. 
\begin{table*}[tbp]
\renewcommand{\arraystretch}{1.3}
\centering
\caption{
Confusion Matrix of the agreement between computer and human judgments.
It shows statistical significance by using  $X^2$ test.
}
\label{tab:count_human_total}
  \begin{tabular}{|c||c|c||c|}
    \hline
    \diagbox{\bfseries human}{\bfseries computer}& \bfseries mono & \bfseries poly & \bfseries total \\
    \hline \hline
    \bfseries mono & 19 & 1 & 20\\ 
    \hline
    \bfseries poly & 1 & 3 & 4\\ 
    \hline \hline
    \bfseries total & 20 & 4 & 24\\
    \hline 
\end{tabular}
\end{table*}
As there exists a case for which the number is less than or equal to $5$, we need Yate's continuity correction.
It achieves statistical significance with level of $\alpha=0.05$.
The disagreement in POLY in Table~\ref{tab:count_human_total} for the word ``james'' attracted our attention.
 
\section{Error analysis}
The disagreement in MONO could be because we chose $3\sigma$, which can detect polysemicity in extremely apparent cases.
Even so, the word ``james'' passes the proposed statistical test.
Therefore, the word ``james'' is worth investing in.

After examining the context of ``james'', we found that it can be used as the name of river and a person.
Table~\ref{tab:name_and_river} shows the various names and how many times the name is used with the word ``river''.
\begin{table*}[tbp]
\renewcommand{\arraystretch}{1.3}
\centering
\caption{
Frequencies of a person's name and the name followed by the word ``river''.
The name``james'' is the most frequently used name with the word ``river''.
}
\label{tab:name_and_river}
  \begin{tabular}{|lr|lr||r|}
      \hline 
	\bfseries name & \bfseries freq & \bfseries river & \bfseries freq & \bfseries ratio (river/name)\\
	\hline
	\hline
	james & 27678 & james river & 202 & 0.007298 \\
	john & 61374 & john river & 76 & 0.001238 \\
	robert & 21491 & robert river & 1 & 0.000047 \\
	richard & 16838 & richard river & 0 & 0.000000 \\
	william & 29310 & william river & 0 & 0.000000 \\
	henry & 19779 & henry river & 0 & 0.000000 \\
	michael & 17854 & michael river & 0 & 0.000000 \\
	charles & 14998 & charles river & 0 & 0.000000 \\
	edward & 23123 & edward river & 112 & 0.004844 \\
	david & 11745 & david river & 0 & 0.000000 \\
	\hline 
\end{tabular}
\end{table*}
The word ``james'' is most frequently used with ``river''.
This may make the word pass the statistical test.
 
\section{Discussion}
The majority of the polysemicity presented in this paper exists due to the Word2Vec compute the distributed representation after ignoring cases. 
This polysemicity might not be regarded as polysemicity with more careful preprocessing. 

The behavior of proposed method depends on the Word2Vec options and the size of the corpus.
If Word2Vec does not have a reasonable neighbor that consists of words of similar usage, the proposed method cannot work effectively.
In addition, a problem arising due the use of Word2Vec for our application is the placement of the vector ``sir'' and the vector ``william'' in similar position.
Therefore, we may need to utilize another method to compute the distributed representation of words.
We use the FIL9 corpus for the experiment.
Though this corpus is freely available to everyone, the size may not be sufficient.
Although we can detect the polysemicity of ``may'', we cannot detect the polysemicity of ``august'' and ``march''.
The statistical test cannot detect the right answer if we do not have sufficient data; therefore, this failure may be interpreted as insufficient usage of ``march'' as verb, and ``august'' as adverb, owing to its origin from Wikipedia, which is in essence a description of facts.

We believe we need to find a way to select the number of neighbors to improve the accuracy of the test.
To make the statistical test more accurate, we need more samples from the neighbors.
At the same time, since we assume that we can measure the statistical fluctuation from the neighbors, we need to exclude words of a different nature from the neighbors.
It is natural that the right number for a neighbor may be different according to the word.
The number that we choose is the minimum value for the statistical test, and has room to adjust for improvement.
  
We computed the neighbor and surrounding uniformity of the 1000 most frequently used words in FIL9.
We observed that proper nouns tend to have a large surrounding uniformity, whereas prepositions tend to have a small surrounding uniformity.
It is an interesting observation that the surrounding uniformity reflects the part of speech information, although it is difficult to determine the class of a word from the value of the surrounding uniformity alone.
For the ease of confirming this observation, the obtained table can be downloaded from the reference (\url{http://www.ss.cs.tut.ac.jp/FIL9SU/}).

\section{Conclusion}
In this paper, we proposed a method to detect polysemy based on the distributed representation by Word2Vec.
We computed the surrounding uniformity of word vector and formed a statistical test.
We illustrated several examples to this measure, and explained the statistical test for detecting polysemy.
In addition, we have also discussed the feasibility of this test.

\bibliographystyle{IEEEtran}
\bibliography{IEEEabrv,kst2017}

\end{document}